# Multi-Stage Prompt-Guided Feature Modulation for Generalizable Brain Tumor Segmentation

Mohammad Mahdi Danesh Pajouh [0009-0009-7039-9233] Sara Saeedi

University of Calgary
Mohammadmahdi.danesh@ucalgary.ca

**Abstract.** Accurate brain tumor segmentation from magnetic resonance imaging (MRI) is essential for diagnosis, treatment planning, surgical guidance, and disease monitoring. However, developing automated segmentation models that generalize across diverse tumor characteristics, imaging protocols, acquisition sites, and patient populations remains challenging. Variations in tumor morphology and imaging distributions can substantially degrade performance outside the training domain. Consequently, improving the robustness and generalization of deep learning-based segmentation models has become a key objective in medical image analysis.
To improve segmentation robustness, we propose Multi-Stage Dynamic Prompt nnU-Net, a prompt-conditioned extension of nnU-Net. Three independent dynamic prompt modules are inserted into the deepest encoder stages. Each module contains a learnable bank of ten 256-dimensional prompt vectors and uses globally pooled encoder features to generate image-specific prompt representations. These representations are projected into feature-wise scaling (γ) and shifting (β) parameters that modulate encoder feature maps through Feature-wise Linear Modulation (FiLM), enabling adaptive feature conditioning at multiple semantic levels.
Evaluation on the BraTS GOAT validation dataset demonstrated that the proposed Multi-Stage Dynamic Prompt nnU-Net outperformed the baseline nnU-Net across the majority of evaluated metrics and tumor subregions. The proposed model achieved average lesion-wise Dice scores of 76.16% (ET), 80.04% (TC), and 86.42% (WT), compared with 74.38%, 78.14% and 84.01% for the baseline model. The results demonstrate that multi-stage dynamic prompt conditioning improves segmentation accuracy and boundary delineation for brain tumor segmentation.

**Keywords:** Brain Tumor Segmentation, nnU-Net, Dynamic Prompting, Domain Generalization

## 1 Introduction

Accurate brain tumor segmentation from magnetic resonance imaging (MRI) is essential for diagnosis, treatment planning, surgical guidance, radiotherapy planning, and longitudinal disease monitoring. In recent years, deep learning methods, particularly U-Net-based architectures, have achieved remarkable performance in automated brain tumor segmentation. Among these approaches, nnU-Net has emerged as a widely adopted baseline due to its self-configuring framework and strong performance across a broad range of medical image segmentation tasks.

Despite these advances, the deployment of segmentation models in real-world clinical scenarios remains challenging due to variations in tumor characteristics and imaging conditions. Brain tumors exhibit substantial differences in size, shape, anatomical location, and appearance, while MRI data can vary across institutions, scanners, acquisition protocols, and patient populations. These variations introduce domain shifts that may significantly reduce model performance when applied to data outside the training distribution. Therefore, developing segmentation models that can adapt to heterogeneous clinical environments remains a major challenge in medical image analysis.

The BraTS challenge has historically provided standardized benchmarks for brain tumor segmentation [1], expanding over time to include distinct tumor populations such as adult gliomas [2, 3], intracranial meningiomas [4], brain metastases [5], and pediatric tumors [6]. In BraTS 2026, these previously separate tumor-specific segmentation challenges are consolidated within the Generalizability Across Tumors (GOAT) task, placing greater emphasis on evaluating algorithmic generalization across diverse clinical domains [7]. GOAT incorporates multiple tumor populations, including adult gliomas, meningiomas, and brain metastases, while validation cohorts introduce additional variability through pediatric tumors and sub-Saharan African glioma populations [8]. This setting reflects realistic sources of distribution shift, including differences in lesion characteristics, demographics, imaging protocols, and the presence or absence of specific tumor subregions. Such evaluations mirror a clinical scenario where segmentation models may encounter previously unseen tumor populations with limited access to additional annotated data.

A potential strategy for addressing such heterogeneity is to enable segmentation networks to dynamically adapt their feature representations according to the characteristics of each input image. Instead of relying on a single fixed representation learned from the training data, adaptive conditioning mechanisms can allow models to adjust their internal responses based on image-specific information. Such approaches are particularly suitable for multi-domain medical imaging problems, where different tumor populations may exhibit distinct imaging patterns and anatomical characteristics.

In this work, we propose Multi-Stage Dynamic Prompt nnU-Net [9], a prompt-conditioned extension of nnU-Net designed to improve robustness across heterogeneous brain tumor segmentation scenarios. Three independent dynamic prompt modules are incorporated into the deepest encoder stages, where they generate image-specific feature modulation parameters through Feature-wise Linear Modulation (FiLM). By adaptively conditioning encoder representations at multiple semantic levels, the proposed framework enables the model to adjust its feature responses according to input-specific characteristics. Experimental evaluation on the BraTS 2026 GOAT validation dataset demonstrates improved performance over the nnU-Net baseline across lesion-wise Dice, Normalized Surface Dice (NSD), and Hausdorff Distance 95 (HD95) metrics for the majority of evaluated tumor subregions.

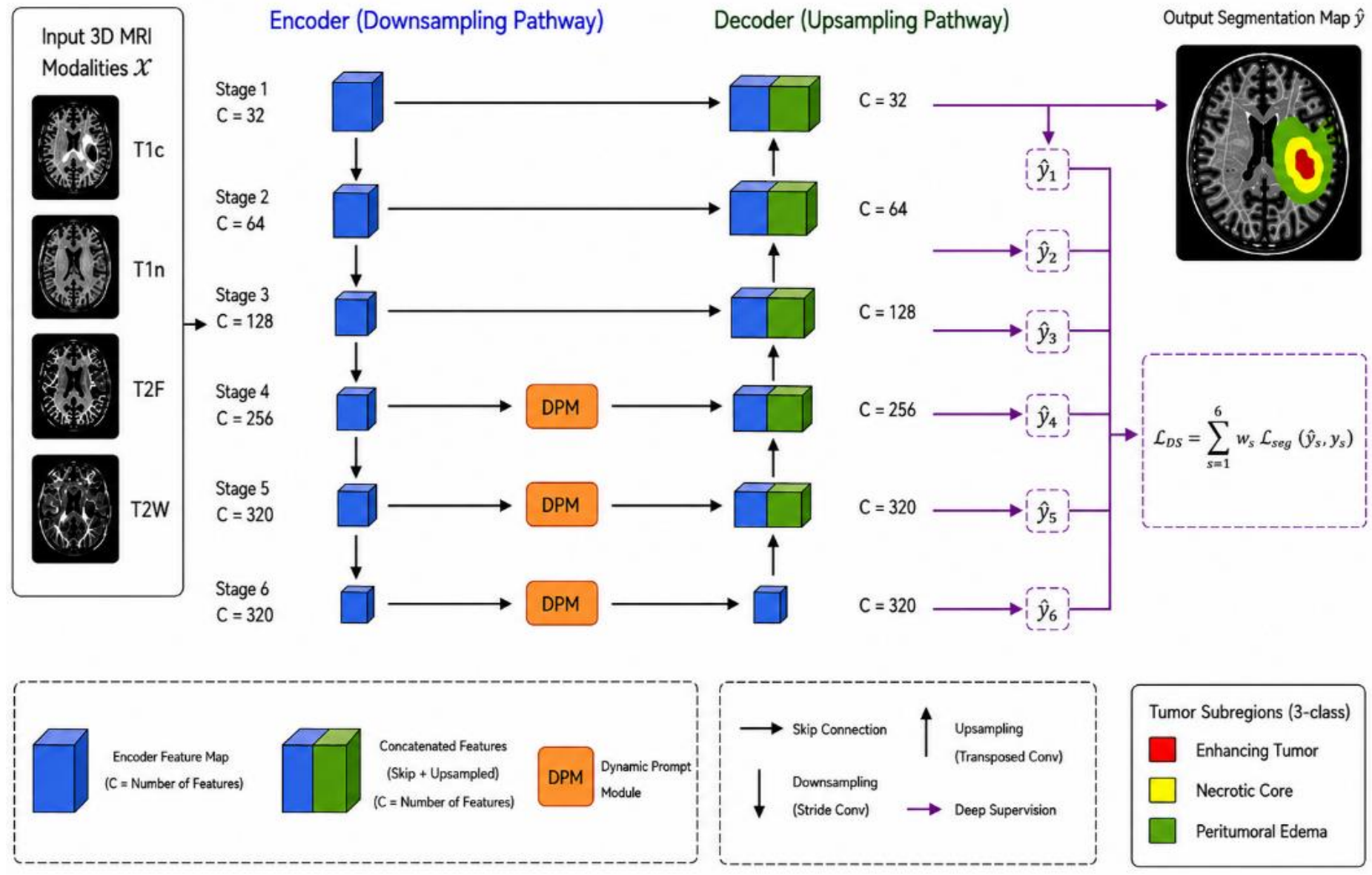

**Fig. 1.** Proposed 3D U-Net architecture. Four MRI modalities are processed through a standard downsampling encoder. Skip connections at the three deepest encoder stages (C = 256, 320, 320) are enhanced using Dynamic Prompt Modules (DPM) prior to concatenation in the upsampling decoder. A deep supervision mechanism computes intermediate losses across all decoder stages to refine multiscale feature learning. The final output delineates three distinct tumor subregions.

## 2 Methodology

### 2.1 Overview of the Dynamic Multi-Prompt Architecture

Medical image segmentation pipelines, particularly those applied to heterogeneous pathologies like brain tumors, traditionally rely on static convolutional kernels. While highly optimized frameworks like nnU-Net [9] establish robust spatial hierarchies, their static nature means the same network weights must accommodate the extreme morphological and textural variations of all tumor sub-regions across all patients.

To overcome this limitation without drastically increasing the computational footprint, we introduce a Dynamic Multi-Prompt framework into the nnU-Net architecture. Instead of relying solely on static feature extraction, our method introduces instance-aware adaptability by injecting input-dependent prompt embeddings. We insert these Dynamic Prompt Modules at the three deepest encoder stages illustrated in Figure 1. At each of these stages, the module evaluates the extracted volumetric features, computes a soft-routing distribution to weigh a bank of learned prompts, and applies a Feature-wise Linear Modulation (FiLM) conditioning step [10]. This allows the network to recalibrate the encoder representations right before they are propagated to the decoder pathway via the skip connections.

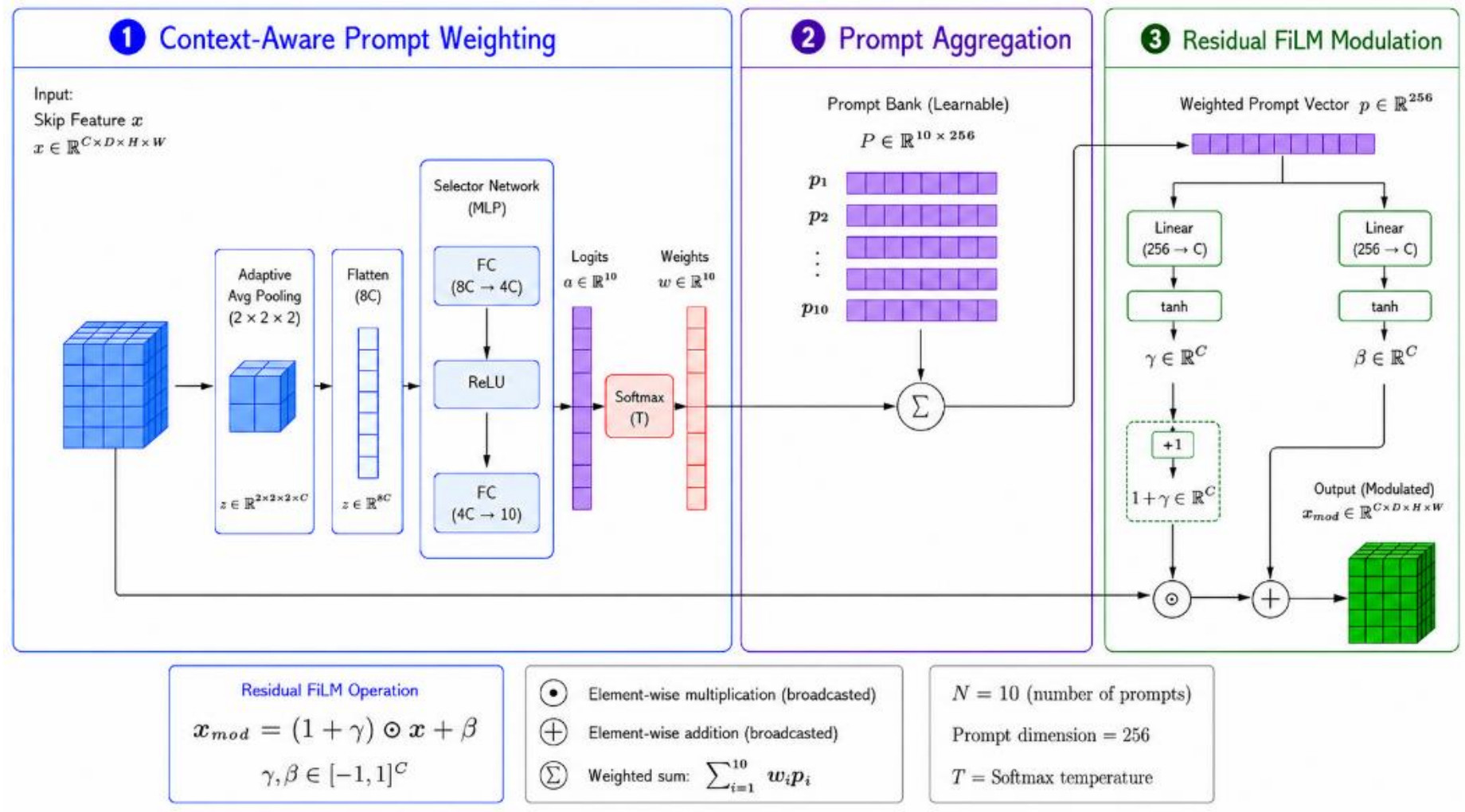

**Fig. 2.** The Dynamic Prompt Module pipeline. The input skip feature x is first pooled and processed by a Selector Network to predict routing weights. These weights are used to aggregate a learnable Prompt Bank into a single context-aware prompt vector. Finally, this vector is projected to scale (γ) and shift (β) parameters to modulate the original feature map via a residual Feature-wise Linear Modulation (FiLM) operation, yielding $x_{mod}$.

## 2.2 Dynamic Multi-Prompt Module

The dynamic prompt module operates sequentially through contextual spatial summarization for prompt weighting, followed by affine feature modulation. The module is depicted in Figure 2.

### Context-Aware Prompt Weighting

To determine which prompts are most relevant for a given input, the network must summarize the feature map $x \in R^{C \times D \times H \times W}$. Standard approaches typically use Global Average Pooling (GAP), which destroys spatial awareness, a highly detrimental effect in brain imaging where anatomical location carries vital semantic clues.

Instead, we employ a 3D adaptive average pooling operation to compress the feature map to a fixed spatial grid of 2 × 2 × 2:

$$z = AdaptivePool_{2 \times 2 \times 2}(x) \tag{1}$$

This preserves coarse-grained spatial awareness by dividing the brain volume into eight distinct spatial octants, resulting in a flattened descriptor vector $z \in R^{8C}$.

This spatially-aware descriptor is passed through a lightweight selector network comprising two fully connected layers interspersed with a ReLU activation to generate selection logits:

$$a = f_{selector}(z) \tag{2}$$

To produce a normalized probability distribution, the logits are processed through a temperature-scaled softmax function:

$$w_i = \frac{e^{\frac{a_i}{T}}}{\sum_j e^{\frac{a_j}{T}}} \quad (3)$$

A temperature-scaled softmax is used to control the sharpness of the prompt-selection distribution. During the initial training phase, a higher temperature (T=0.5) produces a smoother distribution over the prompt bank, ensuring that all prompts receive non-negligible weights and can therefore receive gradient updates. In the later phase, the temperature is reduced to T=0.09, producing sharper prompt assignments and encouraging more specialized selection of the most relevant prompts.. The final aggregated prompt representation p is constructed as a dynamically weighted sum of the entire prompt bank:

$$p = \sum_{i=1}^{N} w_i p_i \quad (4)$$

**Residual Feature-wise Linear Modulation (FiLM)**

The aggregated prompt p interacts with the dense features via Feature-wise Linear Modulation (FiLM) [10]. The prompt p is projected through linear layers to generate channel-wise scaling (γ) and shifting (β) parameters. To constrain the modulation strength and improve training stability, these raw projected parameters are passed through a hyperbolic tangent activation function, bounding their values between -1 and 1.

$$\gamma = \tanh\left(W_\gamma p + b_\gamma\right) \quad (5)$$

$$\beta = \tanh\left(W_\beta p + b_\beta\right) \quad (6)$$

To preserve the integrity of the semantic features established by the base nnU-Net, we employ a residual modulation formulation:

$$x_{modulated} = (1 + \gamma) \odot x + \beta \quad (7)$$

Where $\odot$ denotes element-wise multiplication broadcasted across spatial dimensions. If the linear projections are initialized near zero, γ and β approach zero, making the operation effectively act as an identity mapping: $x_{modulated} \approx x$. This prevents the untrained prompt module from destabilizing the forward pass early in training.

**Multi-Stage Prompt Conditioning**

A separate Dynamic Prompt Module is attached to the skip connections of the three deepest encoder stages. Since skip connections at different depths encode features with varying semantic abstraction, conditioning multiple stages enables the decoder to receive complementary, adaptively modulated representations across multiple semantic scales.

## 2.3 Training Strategy and Loss Formulation

The network is optimized using a combination of region-based segmentation objectives, deep supervision, and orthogonality regularization.

**Segmentation Loss**
To handle class imbalance, we employ a composite loss function combining soft Dice Loss [11] and Focal Loss configured with a focusing parameter of $\gamma = 2$ [12]:

$$L_{seg}(\hat{y}, y) = \lambda L_{Dice} + (1 - \lambda) L_{focal} \tag{8}$$

The loss weighting was also adjusted throughout training. Initially, a larger weight was assigned to the Dice loss $\lambda = 0.6$, with the complementary Focal Loss weight set to $1 - \lambda$, prioritizing high-overlap segmentation during the early stages of training. The Dice weight was then gradually reduced toward a final balanced weighting of $\lambda = 0.5$, allowing the Focal Loss contribution to increase progressively. This schedule was motivated by the observation that the final balanced weighting yielded fewer false-negative predictions in our experiments, consistent with the role of Focal Loss in placing greater emphasis on difficult and misclassified voxels.

**Multi-Scale Deep Supervision**
Deep supervision provided natively by nnU-Net was retained during training. The segmentation loss is evaluated at multiple decoder resolutions, weighted according to the default nnU-Net deep supervision scheme, where weights decay exponentially at lower resolutions and are normalized to sum to one:

$$L_{DS} = \sum_s w_s L_{seg}(\hat{y_s}, y_s) \tag{9}$$

**Orthogonality Regularization**

To prevent prompt collapse, where multiple prompt vectors converge toward similar representations and reduce the effective diversity of the prompt bank, we apply an orthogonality penalty [13]. For each of the M = 3 modulated encoder stages, the L2-normalized prompt matrix $\hat{P}_m$ is constrained to approximate an identity matrix:

$$L_{ortho} = \left(\frac{1}{M}\right) \sum_{m=1}^{M} \left\| \widehat{P_m} \widehat{P_m}^T - I \right\|_F^2 \tag{10}$$

The final objective function is:

$$L_{total} = L_{DS} + \alpha L_{ortho} \tag{11}$$

Where $\alpha = 0.01$ controls the strength of the regularization.

### 2.4 Implementation Details

**Prompt Configuration**
Each Dynamic Prompt Module maintains a distinct bank of N=10 learnable prompt vectors, each with a dimensionality of 256.

**Two-Stage Training Curriculum**
To ensure stable convergence and prevent prompt collapse, training was divided into two distinct phases.

*Phase 1 (Warm-up and Exploration):* In the first half of training, the softmax temperature was set to T=0.5 to encourage a softer routing distribution, allowing gradients to update all prompts uniformly. During this phase, the base nnU-Net backbone, the selector network, and the prompt embeddings were optimized using the same base learning rate of 0.1, allowing the components to co-adapt.

*Phase 2 (Specialization):* In the second half of training, the temperature was annealed to T=0.09 to sharpen the softmax distribution, forcing the network to make confident, specialized prompt selections. Simultaneously, a hierarchical learning rate strategy was introduced: the learning rates for the base U-Net parameters and the selector network were reduced to 0.01× and 0.1× of the base rate, respectively. By constraining the backbone, the dynamic prompt embeddings (which maintained the base learning rate) were forced to learn highly specific and meaningful semantic representations.

**Hardware and Optimization**
The proposed architecture was integrated into the nnU-Net V2 pipeline and was trained for 675 epochs. Experiments were conducted on a single NVIDIA GTX 1080 GPU (8 GB VRAM), utilizing a batch size of 1 and Automatic Mixed Precision (AMP) to maintain numerical stability within memory constraints.

During model development, we also evaluated several alternative configurations, including increasing the prompt bank from 10 to 16 prompts, increasing the prompt dimensionality from 256 to 512, and applying Dynamic Prompt Modules to additional encoder stages. These configurations did not yield consistent improvements on the validation set. Longer training schedules were also explored but did not provide consistent performance gains. Therefore, the final architecture was selected based on the observed validation performance while maintaining a relatively compact prompt configuration and computational complexity.

Optimization was driven by Stochastic Gradient Descent (SGD) utilizing Nesterov momentum and a polynomial decay scheduler. Gradients for the base U-Net were clipped at a maximum norm of 12.0, while prompt-related parameters were clipped at a maximum norm of 1.0.

**Baseline Configuration**
To ensure a fair comparison, the baseline nnU-Net was trained using the same preprocessed images, nnU-Net configuration, and training duration as the proposed model. Both models were trained for 675 epochs with a batch size of 1 using the same 3D full-resolution backbone and input patch configuration. The backbone consists of six encoder stages with feature widths of 32, 64, 128, 256, 320, and 320 channels, using 3D convolutions with InstanceNorm3d normalization and LeakyReLU activation, with two convolutional layers per encoder stage. The 3D full-resolution configuration uses a patch size of 128 × 160 × 112. The baseline uses the default nnU-Net segmentation loss, combining Dice loss and cross-entropy loss, whereas the proposed model additionally incorporates Dynamic Prompt Modules at the three deepest encoder stages and the associated prompt-specific training components.

# 3 Experiments and Results

## 3.1 Quantitative Segmentation Performance

On the validation set, for each subregion namely the enhancing tumor (ET), tumor core (TC) and whole tumor (WT), the average lesion-wise dice, NSD, HD95 and F1 results of baseline nnU-Net and the Dynamic Prompt Modulated (DPM) nnU-Net are presented in Table 1 and Table 2 respectively. Table 3 presents the median lesion-wise Dice and NSD across the three subregions.

**Table 1.** Average Lesion-wise Dice and HD95 of the baseline nnU-Net and DPM nnU-Net on the validation set

| Model | Dice ET (%) | Dice TC (%) | Dice WT (%) | HD95 ET | HD95 TC | HD95 WT |
|---|---|---|---|---|---|---|
| baseline nnU-Net | 74.38 | 78.14 | 84.01 | 46.17 | 22.61 | 19.58 |
| DPM (ours) | 76.16 | 80.04 | 86.42 | 41.18 | 23.46 | 17.01 |

**Table 2.** Average Normalized Surface Dice (NSD) and F1 results of the baseline nnU-Net and DPM nnU-Net on the validation set

| Model | NSD ET (%) | NSD TC (%) | NSD WT (%) | F1 ET (%) | F1 TC (%) | F1 WT (%) |
|---|---|---|---|---|---|---|
| baseline nnU-Net | 49.94 | 46.01 | 42.95 | 61.08 | 68.58 | 66.14 |
| DPM (ours) | 51.08 | 45.82 | 44.31 | 64.54 | 72.19 | 65.93 |

**Table 3.** Median Lesion-wise Dice and NSD of the baseline nnU-Net and DPM nnU-Net on the validation set

| Model | Dice ET (%) | Dice TC (%) | Dice WT (%) | NSD ET (%) | NSD TC (%) | NSD WT (%) |
|---|---|---|---|---|---|---|
| baseline nnU-Net | 88.11 | 92.21 | 92.34 | 53.39 | 49.68 | 41.73 |
| DPM (ours) | 88.66 | 92.08 | 92.78 | 54.12 | 47.7 | 44.17 |

Tables 1–3 compare the proposed Dynamic Prompt Modulated (DPM) nnU-Net with the baseline nnU-Net on the BraTS GOAT validation set. Overall, the proposed model improves most segmentation metrics while introducing only minor degradations for a small number of individual measurements.

The largest improvements are observed for the whole tumor (WT), where the average Dice increases from 84.01% to 86.42%, accompanied by a reduction in HD95 from 19.58 to 17.01. For the enhancing tumor (ET), the proposed method consistently improves Dice, NSD, F1-score, and HD95, indicating more accurate segmentation and better boundary localization. Tumor core (TC) segmentation also benefits from improved Dice and F1-score, although the baseline achieves slightly better HD95 and NSD.

The median metrics further demonstrate that the proposed approach maintains robust performance across the validation cohort. Median Dice improves for ET and WT while remaining comparable for TC. Likewise, median NSD increases for ET and WT, with only a small reduction for TC. These results suggest that the proposed prompt-guided feature modulation improves overall segmentation quality while preserving the strong robustness of the nnU-Net backbone, particularly for the most challenging tumor subregions.

**Training Set Summary.** The final model was trained using the nnU-Net fold_all configuration, in which all available training cases were used for optimization rather than being partitioned into separate training and validation folds. Consequently, model performance during optimization was monitored using the internal training evaluation implemented by nnU-Net. After 675 epochs, the model converged with a training loss of −0.7874 and achieved training pseudo Dice scores of 0.9161, 0.9103, and 0.9409 for the enhancing tumor (ET), tumor core (TC), and whole tumor (WT), respectively.

### 3.2 Computational Complexity

Table 4 summarizes the parameter distribution of the proposed architecture. The baseline nnU-Net backbone contains 30.79 M parameters, while the introduced prompt-related components add approximately 9.16 M parameters, resulting in a total of 39.95 M trainable parameters. Most of the additional parameters originate from the lightweight selector networks, whereas the prompt embeddings themselves require only 7,680 parameters.

Despite the increase in model capacity, inference efficiency is largely preserved. As shown in Table 5, the average inference time increases only from 3.528 s to 3.649 s per volume.

The proposed framework introduces approximately 29.7% more trainable parameters while increasing inference time by only 3.4%, demonstrating that most of the additional computation is confined to lightweight feature modulation rather than expensive convolutional operations. These results indicate that the proposed dynamic prompt mechanism improves segmentation accuracy with only a modest computational cost.

**Table 4.** Parameter count for the different modules used in DPM nnU-Net

| Module | Parameters Count |
|---|---|
| Backbone | 30,790,196 |
| Prompt Vectors | 7,680 |
| Selector Networks | 8,690,206 |
| Gamma/Beta Projectors | 460,544 |
| Total | 39,948,626 |

**Table 5.** Comparison between Average Inference Time of the baseline and DPM nnU-Net

| Model | Average Inference Time |
|---|---|
| baseline nnU-Net | 3.528 s |
| DPM nnU-Net (ours) | 3.649 s |

### 3.3 Prompt Behavior Analysis

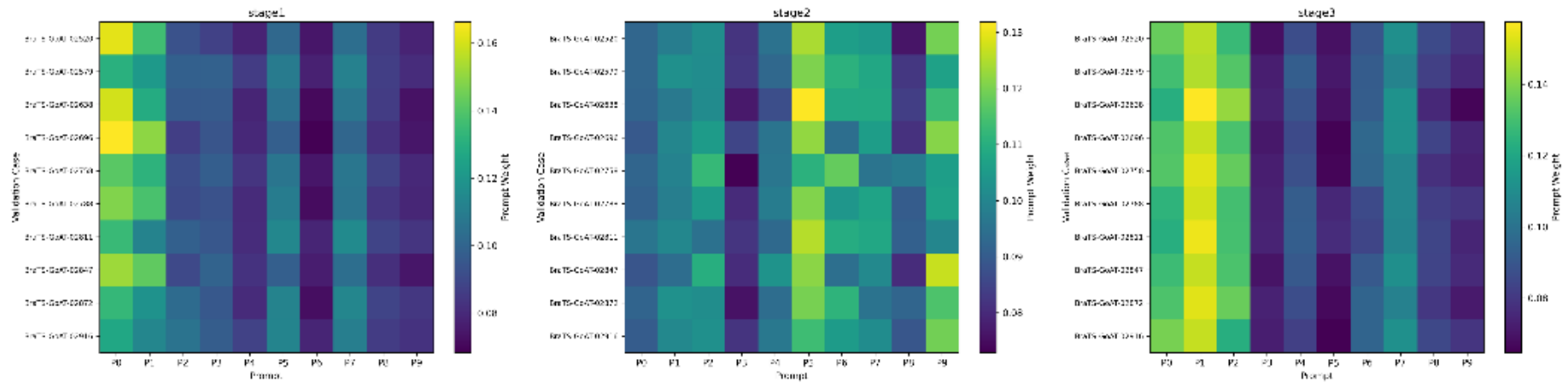

**Fig. 3.** Prompt selection heatmaps for the three Dynamic Prompt Modules. Each row corresponds to a validation subject, and each column represents one learnable prompt. Color intensity represents the average prompt selection weight across all sliding-window patches. The three encoder stages exhibit distinct routing-weight distributions, with varying degrees of prompt utilization across validation subjects.

Figure 3 presents the average prompt selection weights for ten randomly chosen validation cases across the three prompt modules. The learned routing patterns are clearly non-uniform, indicating that the selector network does not distribute the routing weights uniformly across all prompts. Different encoder stages also exhibit distinct prompt utilization patterns, suggesting that the three prompt pools learn different routing behaviors.

The shallow prompt module (Stage 1) demonstrates relatively consistent routing behavior, with Prompt P0 receiving the highest average weight for most validation cases. This observation is consistent with the role of early encoder layers, which primarily encode low-level visual characteristics such as intensity distributions, edges, and local textures. Since these image properties are relatively similar across preprocessed brain MRI scans, a stable routing strategy is expected.

The intermediate prompt module (Stage 2) exhibits noticeably greater variation in prompt selection across the validation cases. Intermediate encoder representations capture increasingly complex structural information, including anatomical context and tumor morphology, which naturally vary more between patients. Consequently, the selector appears to employ a more diverse combination of prompts at this stage, indicating greater case-dependent variation in the routing weights.

In contrast, the deepest prompt module (Stage 3) again displays relatively consistent routing, with Prompt P1 receiving the highest average weight for most cases. Deep encoder features are generally expected to encode more abstract semantic representations than shallow features, which may be associated with less variation in prompt selection across subjects.

Overall, the prompt selection analysis suggests that the proposed multi-stage architecture exhibits different routing behaviors at different representation levels. Rather than converging to identical prompt usage, the three prompt modules exhibit different routing patterns at their respective depths, with relatively stable routing at shallow and deep stages and the greatest case-dependent variation occurring at the intermediate stage.

## 4 Conclusion

In this study, we addressed the critical challenge of domain generalization in automated brain tumor segmentation by introducing the Multi-Stage Dynamic Prompt nnU-Net. Recognizing that static network weights struggle to accommodate the extreme morphological and textural heterogeneity of diverse tumor populations, such as those consolidated in the BraTS GOAT challenge, we enhanced the standard nnU-Net architecture with instance-aware adaptability. By integrating Dynamic Prompt Modules at the deepest encoder stages, our framework leverages spatially-aware feature pooling and a soft-routing mechanism to generate image-specific prompt embeddings. These prompts dynamically recalibrate the extracted semantic features through Feature-wise Linear Modulation (FiLM) prior to decoding, enabling the network to adapt to the unique characteristics of each input volume.

Extensive evaluation on the BraTS GOAT validation dataset demonstrates that the proposed Multi-Stage Dynamic Prompt nnU-Net yields overall improvements in segmentation quality. Improvements are observed across the majority of lesion-wise overlap, boundary, and detection metrics, indicating that dynamic prompt-guided feature modulation effectively enhances feature representations for robust brain tumor segmentation.

Ultimately, this work illustrates that shifting from static feature extraction to adaptive, prompt-guided feature modulation is a highly effective strategy for building robust, generalizable segmentation models capable of navigating the complex, heterogeneous realities of clinical neuro-oncology.

**Acknowledgments.** The authors would like to thank Hossein Danesh Pajouh and Masoumeh F. Mousavi for their encouragement and for providing the equipment used in this research.

Data used in this publication were obtained as part of the Challenge project through Synapse ID (syn74274097).

**Code Availability.** The implementation of the proposed Multi-Stage Dynamic Prompt nnU-Net is publicly available at: https://github.com/LordNecromancer/BraTs-2026-Source-Code.